  \documentclass[onecolumn,usenatbib]{mn2e}
\usepackage{epsf}
\usepackage{epsfig}
\usepackage{amsmath}
\newcommand{\simgt}{\lower.5ex\hbox{$\; \buildrel > \over \sim \;$}}
\newcommand{\simlt}{\lower.5ex\hbox{$\; \buildrel < \over \sim \;$}}

\begin{document}
\title[TeV Cosmic Ray Electrons from Millisecond Pulsars]
{TeV Cosmic Ray Electrons from Millisecond Pulsars}

\author[S. Kisaka and N. Kawanaka]
{Shota Kisaka$^{1 \ast}$,
Norita Kawanaka$^{2 \star}$\\
$^1$ Department of Physics, Hiroshima University, Higashi-Hiroshima 739-8526, Japan \\
$^2$ Racah Institute of Physics, Hebrew University of Jerusalem, Jerusalem, 91904, Israel \\
Email: $^\ast$ kisaka@theo.phys.sci.hiroshima-u.ac.jp, $^\star$ norita@phys.huji.ac.il}


%
\maketitle
\begin{abstract}

Recent $\gamma$-ray observations suggest that the $\gamma$-ray millisecond pulsar (MSP) population is separated into two sub-classes with respect to the pair multiplicity. Here, we calculate the cosmic ray electron/positron spectra from MSPs. Based on the assumption of the equipartition in the pulsar wind region the typical energy of electrons/positrons ejected by a MSP with the pair multiplicity of order unity is $\sim50$ TeV. In this case, we find that a large peak at 10 - 50 TeV energy range would be observed in the cosmic ray electron/positron spectrum. Even if the fraction of pair starved MSPs is 10\%, the large peak would be detectable in the future observations. We also calculate the contribution from MSPs with high pair multiplicity to the electron/positron spectrum. We suggest that if the multiplicity of dominant MSP population is $\sim 10^3$, electrons/positrons from them may contribute to the observed excess from the background electron/positron flux and positron fraction.

\end{abstract}
%
\section{INTRODUCTION}

The {\it Fermi} Gamma-Ray Space Telescope has detected $\gamma$-ray pulsed emissions from more than twenty millisecond pulsars (MSPs) \citep{Ab11}, which have a rotation angular frequency $\Omega\sim 10^3$s$^{-1}$ and a stellar surface magnetic field $B_s\sim 10^{8.5}$G. The detection of the GeV emissions from a pulsar magnetosphere means that electrons and positrons are accelerated to more than $\sim$ TeV by the electric field parallel to the magnetic field, which arises in a depleted region of the Goldreich-Julian (GJ) charge density \citep{GJ69}. The $\gamma$-ray light curve is an important tool for probing the particle acceleration process in the pulsar magnetosphere. Therefore, the $\gamma$-ray emission region has been explored by comparing theoretical models such as polar cap \citep{DH96}, outer gap \citep{CHR86} and slot gap models \citep{MH04} with the observed light curve (e.g., Venter, Harding \& Guillemot 2009; Romani \& Watters 2010; Kisaka \& Kojima 2011).

\citet{VHG09} fitted the pulse profiles of the {\it Fermi} detected MSPs with the geometries of $\gamma$-ray emission region predicted by different theoretical models. They found that the pulse profiles of six of eight MSPs could be fitted by the geometries of either the outer gap or the slot gap model, as was the case of canonical pulsars. They interpreted that copious pairs are produced in the magnetosphere of these MSPs.  However, \citet{VHG09} also found that the pulse profiles of remaining two MSPs show the unusual behavior in the $\gamma$-ray light curves and could not be fitted by the geometry of either the outer gap or the slot gap models. They proposed that these unusual light curves could be fitted by the pair starved polar cap model \citep{MH04b}, in which the multiplicity of the pairs is not high enough to completely screen the electric field above the polar cap, and the particles are continuously accelerated up to high altitude over the entire open field line region. Thus, from the model fitting of the $\gamma$-ray light curves, \citet{VHG09} suggested that the $\gamma$-ray MSP population is separated into two sub-classes.

The important fact is that radio pulsed emission is also detected from all currently detected $\gamma$-ray MSPs and remarkably similar to that from canonical pulsars. The pulsar radio emission is a highly coherent process because the brightness temperature is extremely high. In the theoretical models of the radio emission mechanisms, some authors have believed the conditions that there are a highly relativistic primary beam with the large Lorentz factor ($\sim 10^7$) and the number density nearly equal to the GJ density, and the secondary electron/positron plasma with relatively small bulk streaming Lorentz factor ($\sim 10$ - $10^3$) and the large pair multiplicity ($\sim 10^3$ - $10^5$) in the radio emitting region (e.g., Melrose 1995; Lyutikov, Blandford \& Machabeli 1999; Gedalin, Gruman \& Melrose 2002). However, the existence of pair starved MSPs suggests that the radio emission mechanisms should be insensitive to the particle number density down to sub-GJ number density. The pulsar radio emission mechanism is still poorly understood, so that the observationally-based constraints are valuable \citep{M95}. Therefore, another verification for the extent of the MSP multiplicity, especially the existence of the pair starved MSPs is important for the pulsar radio emission mechanisms.

Recently, HESS has discovered a new TeV source, which is located in the close vicinity of the globular cluster Terzan 5 \citep{HE11}. Several globular clusters, including Terzan 5 also emit GeV $\gamma$-ray \citep{Ab09b, KHC10, Ab10b, Ta11}, which may plausibly be due to a number of MSPs residing in these clusters \citep{HUM05, VDC09}. Thus, inverse Compton scattering by the high-energy particles ejected from MSPs are proposed for the origin of the observed TeV emission \citep{BS07, VDC09}. The high-energy electron/positron spectrum ejected from MSPs would be a useful probe for the multiplicity of the MSPs. However, only from the TeV spectra, we cannot distinguish two models \citep{BS07, VDC09}, which assume different pair multiplicities \citep{HE11}. Another way to investigate the electron/positron spectrum ejected from MSPs is its direct measurement. Since high-energy electrons/positrons can propagate only about a few kpc due to the energy losses by the synchrotron and the inverse Compton emission, the direct detection of the electrons/positrons ejected from MSPs in the globular clusters is unlikely. However, for the following reasons, we may detect those from nearby MSPs.

MSPs have much lower spin-down luminosity than canonical pulsars. \citet{BVD08} investigated the possible contribution of the nearby MSP, PSR J0437-4715 to the cosmic ray electron/positron spectrum. They concluded that unlike canonical pulsars such as Geminga pulsar, the contribution from a MSP to the observed electron/positron flux is negligible. However, since the lifetime of MSPs is much longer than canonical pulsars ($>10^{10}$ yr), there should be much more nearby active MSPs. Furthermore, Kashiyama, Ioka \& Kawanaka (2011; hereafter KIK11) pointed out that since white dwarf pulsars have long lifetime and continue to inject the electrons/positrons after the nebulae stop expanding, the adiabatic energy losses of electrons/positrons in the pulsar wind nebula region are negligible. Also the synchrotron cooling of electrons/positrons is so small and the high-energy electrons/positrons can escape the nebulae without losing much energy. Although they consider the case of the white dwarfs, their results are also applicable to MSPs. Therefore MSPs could potentially contribute to the observed high energy cosmic ray electrons/positrons and will be detectable by the next generation experiments, such as CALET \citep{To08} and CTA \citep{CTA}.

In this paper, we investigate the contribution of electrons/positrons ejected from the MSPs to the observed cosmic ray spectrum. In section 2, we apply KIK11 model to the case of MSPs. We estimate the typical energy of electrons/positrons from the MSPs and show that during the propagation in pulsar wind nebulae, the adiabatic losses and radiative cooling of electrons/positrons are not so large. We also describe the propagation in the interstellar medium (ISM). In section 3, we calculate the energy spectrum of cosmic ray electrons/positrons from the MSPs and show the possibility that the electrons/positrons from these MSPs are detectable for the future observations. 

\section{THE MODEL}

\subsection{Acceleration and cooling}

In order to estimate the energy of electrons/positrons available in the wind region and their adiabatic and radiative cooling in the shocked region, we adopt the model of KIK11. For a pulsar wind nebula formed by a MSP, we consider the conditions that the relativistic wind blasts off from the light cylinder $\sim R_{\rm lc}=c/\Omega$ where $c$ is the speed of light, and two shock fronts are formed between the supersonic pulsar wind and ISM. Since the energy from a MSP is continuously transported to the wind, the shock fronts are expanding until the pressure of the shocked region $P_{\rm sh}$ becomes equal to that of ISM $p$. Although KIK11 considered the case of white dwarf pulsars, the situations are similar to the case of MSPs because they have a long lifetime and the supernova shock front have already decayed. We assume that the effects of binary companion are negligible, because the fraction of solid angle occupied by companion is small ($<$1\%). We also neglect radiative loss due to curvature radiation within light cylinder ($\sim 10$\%). From now on we set fiducial parameters of the MSP's surface magnetic field strength, angular frequency and radius as $B_0 = 10^{8.5}$G, $\Omega = 10^3$s$^{-1}$ and $R = 10^6$cm, respectively.

We assume the energy equipartition between particles and magnetic field, $\varepsilon_eN=B^2/8\pi$, and the conservation of the particle number flux, $4\pi r^2cN\sim$ constant, in the MSP wind region. Here, $N$ is the number density of electrons/positrons. The number density can be described as 
\begin{equation}
N=N_{\rm lc}\left(\frac{R_{\rm lc}}{r}\right)^2=\frac{B_{\rm lc}\Omega\kappa}{2\pi ce}\left(\frac{R_{\rm lc}}{r}\right)^2,
\end{equation}
where $\kappa$ is the multiplicity of electrons/positrons, $N_{\rm lc}$ and $B_{\rm lc}$ are the number density and the magnetic field at the light cylinder, respectively. We assume the magnetic field configuration as pure dipole ($B\propto r^{-3}$) within light cylinder. Outside the light cylinder, we assume the conservation of the energy flux of the magnetic field, $Br\sim$ constant. Using these assumptions, the typical energy of electrons/positrons $\varepsilon_e$ can be described as
\begin{equation}
\varepsilon_e = \frac{e\psi_{\max}}{\kappa} \sim 50\kappa^{-1}\left(\frac{B_0}{10^{8.5}{\rm G}}\right)\left(\frac{\Omega}{10^3{\rm s^{-1}}}\right)^2\left(\frac{R}{10^{6}{\rm cm}}\right)^3 {\rm TeV},
\label{eq:2.1}
\end{equation}
where $\psi_{\rm max}$ is the electric potential difference across the open magnetic field lines described as 
\begin{equation}
\psi_{\max}=\frac{B_0\Omega^2R^3}{2c^2}\sim 5\times10^{13}\left(\frac{B_0}{10^{8.5}{\rm G}}\right)\left(\frac{\Omega}{10^3{\rm s^{-1}}}\right)^2\left(\frac{R}{10^{6}{\rm cm}}\right)^3 {\rm Volt}.
\label{eq:2.2}
\end{equation}
These values are similar to those in the case of white dwarf pulsars (KIK11). Note that the typical energy depends on the pair multiplicity. 

Next, we estimate the adiabatic and the radiative cooling of electrons/positrons in the shocked region. The outer shock of the pulsar wind nebula finally decays when the pressure of the shocked region $P_{\rm sh}$ becomes equal to that of the ISM $p$. If the outer shock decaying time is shorter than the lifetime of MSP, the adiabatic cooling is negligible. In order to estimate the outer shock decaying timescale, we solve the equation of motion and the energy conservation law at the outer shock front. The equation of motion can be described as 
\begin{equation}\label{eos}
\frac{d}{dt} \left\{ \frac{4 \pi}{3} R_{\rm out}^3 \rho \frac{dR_{\rm out}}{dt} \right\} = 4\pi R_{\rm out}^2 P_{\rm sh},
\end{equation}
where $R_{\rm out}$ is the radius of the outer shock front, $P_{\rm sh}$ is the pressure of the shocked region and $\rho$ is the density of ISM. The energy equation is 
\begin{equation}\label{ece}
\frac{d}{dt} \left\{ \frac{4 \pi}{3} R_{\rm out}^3 \frac{3}{2} P_{\rm sh} \right\} = L_{\rm sd} - P_{\rm sh} \frac{d}{dt} \left\{ \frac{4 \pi}{3} R_{\rm out}^3 \right\} ,
\end{equation}
where $L_{\rm sd}$ is the spin-down luminosity of MSP, 
\begin{equation}
L_{\rm sd} =\frac{B^2_0\Omega^4R^6}{2c^3} = 2\times 10^{33}\left(\frac{B_0}{10^{8.5}{\rm G}}\right)^2\left(\frac{\Omega}{10^3{\rm s^{-1}}}\right)^4\left(\frac{R}{10^{6}{\rm cm}}\right)^6 {\rm erg}\ {\rm s}^{-1}.
\end{equation}
In the derivation of eq.(\ref{ece}), we assume that in the shocked region the internal energy of particles is $3P_{\rm sh}/2$ because the energy of particles is the relativistic regime. Using the typical value for the density of ISM $\rho\sim 10^{-24}{\rm g}\ {\rm cm}^{-3}$, the pressure in ISM is $p\sim 10^{-13} {\rm dyn}\ {\rm cm}^{-2}$. Solving above equations, the outer shock decays at about
\begin{equation}
t_{\rm dec}\sim 10^6 \left(\frac{B_0}{10^{8.5}{\rm G}}\right)\left(\frac{\Omega}{10^3{\rm s^{-1}}}\right)^2\left(\frac{R}{10^{6}{\rm cm}}\right)^3 \left(\frac{T}{10^{3}{\rm K}}\right)^{-5/4} {\rm yr}, 
\label{eq:2.3}
\end{equation}
where $T$ is the temperature of ISM. The lifetime of a MSP $\tau$ can be estimated as 
\begin{equation}
\tau=\frac{E_{\rm rot}}{L_{\rm sd}}, 
\label{eq:2.4}
\end{equation}
where $E_{\rm rot}$ is the rotation energy described as 
\begin{equation}
E_{\rm rot}\sim 10^{52}\left(\frac{M}{1.4M_{\odot}}\right)\left(\frac{R}{10^6{\rm cm}}\right)^2\left(\frac{\Omega}{10^3{\rm s^{-1}}}\right)^2 {\rm erg},
\label{eq:2.5}
\end {equation}
where $M$ is the mass of a MSP. For the fiducial parameters of MSP
\begin{equation}
\tau\sim 5\times 10^{10} \left(\frac{M}{1.4M_{\odot}}\right)\left(\frac{B_0}{10^{8.5}{\rm G}}\right)^{-2}\left(\frac{\Omega}{10^3{\rm s^{-1}}}\right)^{-2}\left(\frac{R}{10^{6}{\rm cm}}\right)^{-4} {\rm yr}.
\label{eq:2.7}
\end{equation}
We find that the outer shock decays at a very early stage of the lifetime of MSP and does not expand after $t > t_{\rm dec}$. Therefore, in the similar to the case of white dwarf pulsars (KIK11), the adiabatic cooling shall give minor contributions to the cooling process of the high-energy electrons/positrons. 

Also for the radiative cooling by the synchrotron radiation in the shocked region $R_{\rm in} < r < R_{\rm out}$, we follow the discussion in KIK11 based on the diffusion in the shocked region. We take the Bohm limit, where the fluctuation of the magnetic field is comparable to the coherent magnetic field strength. In this limit, the timescale $t_{\rm diff}$ for the electron/positron trapping in the shocked region is given by 
\begin{equation}
t_{\rm diff}=\frac{d^2}{2D_{\rm sh}}=\frac{3}{2}\frac{eBd^2}{\varepsilon_ec},
\end{equation}
where $D_{\rm sh}=cr_{\rm g}/3$ is the diffusion coefficient under the Bohm limit, $d$ is the size of the shocked region and $r_g=\varepsilon_e/eB$ is the Larmor radius of electron/positron with energy $\varepsilon_e$. To estimate the diffusion timescale, we need to know the size and the magnetic field strength of the shocked region. Here we consider the time $t>t_{\rm dec}$, so that the size of the shocked region is an order of the radius of forward shock front at $t=t_{\rm dec}$, $d\sim R_{\rm out}(t=t_{\rm dec})\sim 3\times 10^{19}{\rm cm}$. For the radius of the inner shock front at $t>t_{\rm dec}$, we use the balance condition between the momentum transferred by wind and the pressure of ISM
\begin{equation}
\frac{L_{\rm sd}}{4\pi R^2_{\rm in}c}=p.
\end{equation}
For the fiducial parameters, $R_{\rm in}(t>t_{\rm dec})\sim 2\times 10^{17}$ cm.
The strength of the magnetic field at the inner radius $B_{\rm in}$ can be estimated as $B_{\rm in}\sim 2 \times 10^{-6}$ G, which is almost the same as that of ISM. Then, the diffusion timescale is 
\begin{equation} 
t_{\rm diff}\sim 2 \times 10^4 \left(\frac{\varepsilon_e}{50 {\rm TeV}}\right)^{-1} {\rm yr}.
\end{equation}
The synchrotron energy loss of a particle with energy $\varepsilon_e$ is described as 
\begin{equation}
\frac{d\varepsilon_e}{dt}=-\frac{4}{3}\sigma_{\rm T}c\beta^2\frac{B^2}{8\pi}\left(\frac{\varepsilon_e}{m_ec^2}\right)^2,
\end{equation}
where $\sigma_{\rm T}$ is the Thomson scattering cross section, $\beta$ is the particle velocity normalized by the speed of light and $m_e$ is the mass of electron/positron. Then, The typical energy loss of the electrons/positrons $\Delta\varepsilon_e$ with energy $\varepsilon_e$ can be estimated as
\begin{equation}
\frac{\Delta\varepsilon_e}{\varepsilon_e}\sim 0.3\left(\frac{B_0}{10^{8.5}{\rm G}}\right)^2 \left(\frac{\Omega}{10^3{\rm s}^{-1}}\right)^4\left(\frac{R}{10^6{\rm cm}}\right)^6.
\label{eq:2.8}
\end{equation}
This means that the high-energy electrons/positrons injected into the shocked region lose roughly 30\% of the energy by the synchrotron radiation before diffusing out into ISM. Therefore, as in the case of white dwarf pulsars (KIK11), we can conclude that the radiative energy loss of electrons/positrons in the pulsar wind nebula is not so large.

The above expressions for the estimate of the energy losses are only applicable to the case that the velocity of a MSP is subsonic in ISM. The observed velocity of MSPs is less than that of canonical pulsars in average sense \citep{Ho05}. However, some MSPs have the large velocity and a few MSPs actually forms bow shock nebulae \citep{St03, HB06}. In this case, the size of the bow shock is described as (e.g., Wilkin 1996)
\begin{equation}
R_{\rm bow}=\left(\frac{L_{\rm sd}}{4\pi c\rho V^2}\right)^{1/2}\sim 10^{16}\left(\frac{B_0}{10^{8.5}{\rm G}}\right)\left(\frac{\Omega}{10^3{\rm s^{-1}}}\right)^{2}\left(\frac{R}{10^{6}{\rm cm}}\right)^{3}\left(\frac{V}{10^7{\rm cm\ s^{-1}}}\right)^{-1} {\rm cm},
\end{equation}
where $V$ is the velocity of a MSP. Due to the assumption of the energy equipartition, the strength of the magnetic field is 
\begin{equation}
B_{\rm bow}=\left(\frac{2L_{\rm sd}}{cR^2_{\rm bow}}\right)^{1/2}\sim 50\left(\frac{V}{10^7{\rm cm\ s^{-1}}}\right) \mu {\rm G}.
\end{equation}
The ratio of the Larmor radius of electrons/positrons to the bow shock radius is
\begin{equation}
\frac{r_{\rm g}}{R_{\rm bow}}\sim 0.5 \kappa^{-1}.
\end{equation}
The fact that $r_{\rm g}/R_{\rm bow}$ is close to unity supports that electrons/positrons may escape from the bow shock region. Therefore, in the case of $\kappa\sim 1$, high-energy electrons/positrons can escape with an efficiency of order unity \citep{B08, BA10}. Even if we consider the case of $\kappa \gg 1$, the synchrotron loss can be estimated by using eqs. (11), (14) and (16) as
\begin{equation}
\frac{\Delta\varepsilon_e}{\varepsilon_e}\sim 9\times 10^{-4}\left(\frac{B_0}{10^{8.5}{\rm G}}\right)^2 \left(\frac{\Omega}{10^3{\rm s}^{-1}}\right)^4\left(\frac{R}{10^6{\rm cm}}\right)^6\left(\frac{V}{10^7{\rm cm\ s^{-1}}}\right).
\end{equation}
Therefore, we can conclude again that the radiative energy loss of electrons/positrons in the pulsar wind nebula is not so large.

\subsection{Diffusion in Interstellar medium}

The observed electron/positron spectrum after the propagation in ISM is obtained by solving the diffusion equation
\begin{equation}
\frac{\partial}{\partial t}f(t,r,\varepsilon_e)=D(\varepsilon_e)\nabla^2 f+\frac{\partial}{\partial \varepsilon_e}\left( P(\varepsilon_e)f \right) +Q(t,r,\varepsilon_e),
\end{equation}
where $f(t,r,\varepsilon_e)$ is the energy distribution function of electrons/positrons, $D(\varepsilon _e)=D_0(1+\varepsilon_e/3{\rm GeV})^{\delta}$ is the diffusion coefficient, $P(\varepsilon_e)$ is the cooling function of the electrons/positrons which takes into account synchrotron emissions and inverse Compton scatterings during the propagation, and $Q(t,\varepsilon_e,r)$ is the injection term.  Here we adopt $D_0=5.8\times 10^{28}{\rm cm}^2{\rm s}^{-1}$, $\delta=1/3$, which is consistent with the boron-to-carbon ratio according to the latest GALPROP code.  Atoyan et al. (1995) showed a solution in the case of an instantaneous injection from a single point-like source, i.e. $Q(t, \varepsilon_e,r)  \approx Q_0(\varepsilon_e) \delta(t-t_i)\delta(r)$.  Then the observed spectrum $G(t,r,\varepsilon_e; \tilde{t})$ would be
\begin{equation}
G(t,r,\varepsilon_e; \tilde{t})=\frac{Q_0(\varepsilon_{e,0})P(\varepsilon_{e,0})}{\pi ^{3/2} P(\varepsilon_e)d_{\rm diff}(\varepsilon_e, \varepsilon_{e,0})^3}\exp \left(-\frac{r^2}{d_{\rm diff}(\varepsilon_e, \varepsilon_{e,0})^2} \right),
\end{equation}
where $\varepsilon_{e,0}$ is the energy of electrons/positrons at the time $\tilde{t} (<t)$ and which are cooled down to $\varepsilon_e$ at the time $t$, and $d_{\rm diff}$ is the diffusion length given by
\begin{equation}
d_{\rm diff}=2\left[ \int_{\varepsilon_e}^{\varepsilon_{e,0}} \frac{D(x)dx}{P(x)} \right]^{-1/2}. \label{d_diff}
\end{equation}

The cooling function $P(\varepsilon_e)$ can be described as
\begin{equation}
P(\varepsilon_e)=\frac{4\sigma_T \varepsilon_e^2}{3m_e^2 c^3} \left[ \frac{B^2}{8\pi} + \int d\varepsilon_{\gamma} u_{\rm tot} (\varepsilon_{\gamma}) f_{\rm KN} \left( \frac{4\varepsilon_e \varepsilon_{\gamma}}{m_e^2 c^4} \right) \right]
\end{equation}
where $u_{\rm tot}(\varepsilon_{\gamma})d\varepsilon_{\gamma}$ is the energy density of interstellar radiation fields with the photon energy between $\varepsilon_{\gamma}$ and $\varepsilon_{\gamma}+d\varepsilon_{\gamma}$ (including cosmic microwave background, starlight, and dust emission; Porter et al. 2008), and $B$ is the interstellar magnetic field with we here set as $1\mu {\rm G}$.  Here the function $f_{\rm KN}(x)$ is the correction factor to include the Klein-Nishina (KN) effect, which approaches to unity when $x$ is much smaller than unity.  The mathematical expression of $f_{\rm KN}$ can be found in Moderski et al. (2005).

As shown in the last section, in general MSPs have a very long lifetime ($\tau \sim 5\times 10^{10}{\rm yr}$) which is comparable with the cosmic age.  In such a case that a point-like source with a finite duration, taking into account the time-dependence of an injection rate ($Q_0(\varepsilon_e)\rightarrow Q_0(\varepsilon_e, \tilde{t})$), the spectrum can be calculated by integrating $G(t,r,\varepsilon_e;\tau)$ for $\tau$:

\begin{equation}
f_1(t,r,\varepsilon_e; t_i)=\int_{t_i}^{t} d\tilde{t}~G\left( t,r,\varepsilon_e; \tilde{t} \right) ,
\end{equation}
where $t_i$ is the time when the particle injection from a source has started.  Here we assume that the electron/positron injection spectrum can be described as mono-energetic distribution
\begin{equation}\label{mono}
Q_0(\varepsilon_e,\tilde{t})\propto \left( 1+\frac{\tilde{t}-t_i}{\tau} \right)^{-2},
\end{equation}
or power-law distribution
\begin{equation}\label{pl}
Q_0(\varepsilon_e,\tilde{t})\propto \varepsilon_e^{-\alpha}\exp \left( -\frac{\varepsilon_e}{\varepsilon_{\rm cut}}\right) \left( 1+\frac{\tilde{t}-t_i}{\tau} \right)^{-2},
\end{equation}
where $\alpha$ is the intrinsic power-law index of an electron/positron spectrum and $\varepsilon_{\rm cut}$ is the maximum electron/positron energy from a source.

Now we can calculate the average electron/positron spectrum by considering the birth rate of MSPs  as follows:

\begin{equation}
f_{\rm ave}(\varepsilon_e)=\int_0^{t_0} dt_i \int_0^{d_{\rm diff}(\varepsilon_e,\varepsilon_{e,i})} 2\pi r dr f_1(t_0,r,\varepsilon_e; t_i)R,
\end{equation}
where $t_0$ is the cosmic age (i.e. the present time),  and $R$ is the local pulsar birth rate (${\rm yr}^{-1}~{\rm kpc}^{-2}$).  Here $\varepsilon_{e,i}$ is the energy at the time $t_i$ of CR electrons/positrons which are cooled down to $\varepsilon_e$ at $t$.  

\section{RESULTS AND DISCUSSIONS}

First, we calculate the cosmic ray electron/positron spectra from the pair starved MSPs. We set the pair multiplicity $\kappa= 1$, the lifetime $\tau= 5\times 10^{10}$yr, the total energy $E_{\rm rot}= 10^{52}$ erg, the local birth rate $R = 3\times 10^{-9}$ yr$^{-1}$ kpc$^{-2}$ and the fraction of the lost energy due to synchrotron emission 30\% for each MSP. We assume that each MSP has the same value of the parameters ($B_0=10^{8.5}{\rm G}, \Omega=10^3{\rm s}^{-1}, R=10^6{\rm cm}$), because most MSPs have the almost same spin-down luminosity. For the injection distribution function, we assume mono-energetic distribution eq.(\ref{mono}) with the energy $\varepsilon_e=50$ TeV. Even if we consider the power-law distribution eq.(\ref{pl}), the energy range of the distribution is small because the allowed range of the cutoff energy should be only 50 - 80 TeV due to the observed constraint by KASKADE/GRAPES/CASA-MIA \citep{KY09}. Thus the distribution should be nearly mono-energetic distribution. Note that our local MSP birth rate is based on the MSP local surface density of 38$\pm$ 16 pulsars kpc$^{-2}$ for 430 MHz luminosity above 1 mJy kpc$^2$ \citep{L08}. Actually, now 23 MSPs are detected within 1kpc \citep{ATNF}. We can only detect MSPs that have the radio beam directed toward us and the radio flux larger than the threshold of detectors. However, the cosmic ray electrons/positrons ejected from MSPs are distributed isotropically because of the effect of Galactic magnetic field, so that a large number of MSPs will contribute to the observed electron/positron spectrum. Therefore, our local MSP birth rate corresponds to the lower limit for the current radio observations. 

\begin{figure}
 \begin{center}
  \includegraphics[width=160mm]{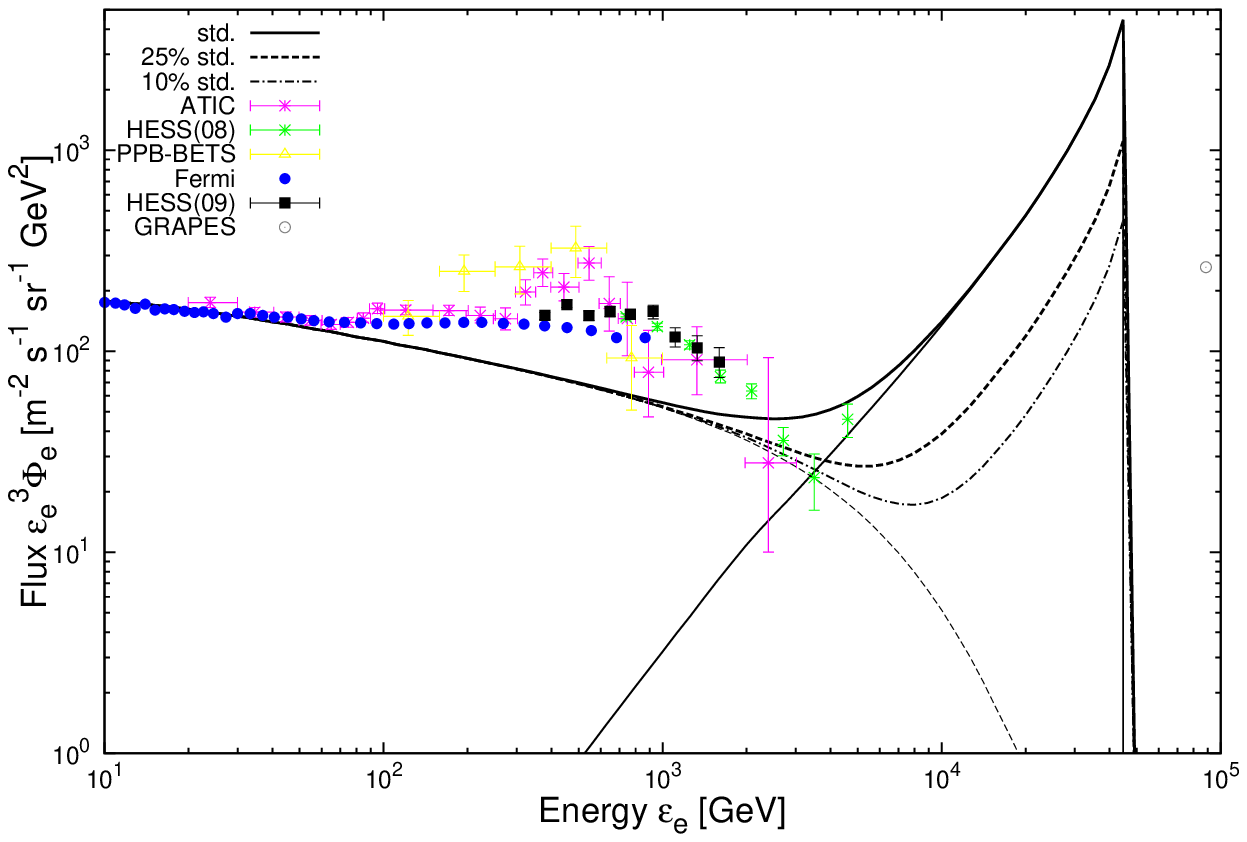}
 \end{center}
 \caption{Electron/positron spectrum predicted from MSPs with the fraction of pair starved MSPs 100\% (thin solid line), and its sum (thick solid line) with the background (thin dashed line). The injection distribution function is the mono-energetic distribution eq.(\ref{mono}) with the energy $\varepsilon_e=50$ TeV and the multiplicity $\kappa = 1$. Data points correspond to measurements of ATIC (purple boxes, Chang et al. 2008), HESS (light-green triangles and black triangles, Aharonian et al. 2008; 2009), PPB-BETS (yellow triangles, Torii et al. 2008b), {\it Fermi} (blue circles, Ackermann et al. 2010) and GRAPES (black circle, Kistler \& Y\"{u}ksel 2009). We also show the total spectra from pair starved MSPs with the different fractions: 25\% (thick dashed line) and 10\% (dot-dashed line). We assume that the lifetime $\tau=5\times 10^{10}$ yr, the total energy $E_{\rm rot}=10^{52}$ erg, the local birth rate $R=3\times 10^{-9}$ yr$^{-1}$ kpc$^{-2}$ and the fraction of the energy loss is 30\%.} 
 \label{fig:3.1}
\end{figure}

In figure \ref{fig:3.1}, the electron/positron flux from multiple pair starved MSPs is shown. Thick solid, dashed and dash-dotted lines show the total electron/positron spectra if the fraction of the pair starved MSPs is 100\%, 25\% and 10\%, respectively. Thin solid line shows the contribution of electrons/positrons from the pair starved MSPs when the fraction is 100\%. The background flux is shown as a thin dashed line. We adopt the background model of an exponentially cutoff power law with an index of -3.0 and a cutoff at 1.5 TeV, which is similar to that shown in \citet{Ah08} and reproduces the data in $\sim$10 GeV-1 TeV well. It is very interesting that there is a large peak at 10-50 TeV energy range. The existence of this peak cannot be ruled out from the current observations. The high-energy component is more enhanced for the long-duration sources \citep{AAV95, KIN10}. This is because the longer the duration of injection is, the larger fraction of fresh electrons/positrons are expected to reach the Earth without losing their energy so much during the propagation. MSPs are the continuously injecting sources with the duration as long as the cosmic age, so that the spectrum from them has nearly the same shape with the injection spectrum with the soft energy tail component. This is the difference from other sources such as young pulsars, whose typical duration is only $\sim 10^4$-$10^5$ yrs. Fitting result of \citet{VHG09} showed that the fraction of the pair starved MSPs is 25\%, although they have only eight samples. Even if the fraction is 10\%, the flux is $\sim 20$m$^{-2}$ s$^{-1}$ sr$^{-1}$ GeV$^2$ at 10 TeV. In this case, we can detect the electron/positron flux with near future missions such as CALET (we assume the geometrical factor times the observation time $\sim 5$ yrs as $\sim 220$ m$^2$ sr days) because the predicted electron/positron flux is sufficiently large. It was considered that the number of astrophysical sources contributing to the above several TeV energy range is quite small according to the birth rate of the supernovae and the canonical pulsars in the vicinity of the Earth \citep{Ko04, KIN10, KIOK11}. However, we find that it is possible for multiple pair starved MSPs to contribute to the 10 TeV energy range in the electron/positron spectrum. Therefore, if the anisotropy of the observed electrons/positrons are weak in the 10 TeV energy range, we suggest that the pair starved MSPs may contribute to the spectrum significantly. 

\begin{figure}
 \begin{center}
  \includegraphics[width=160mm]{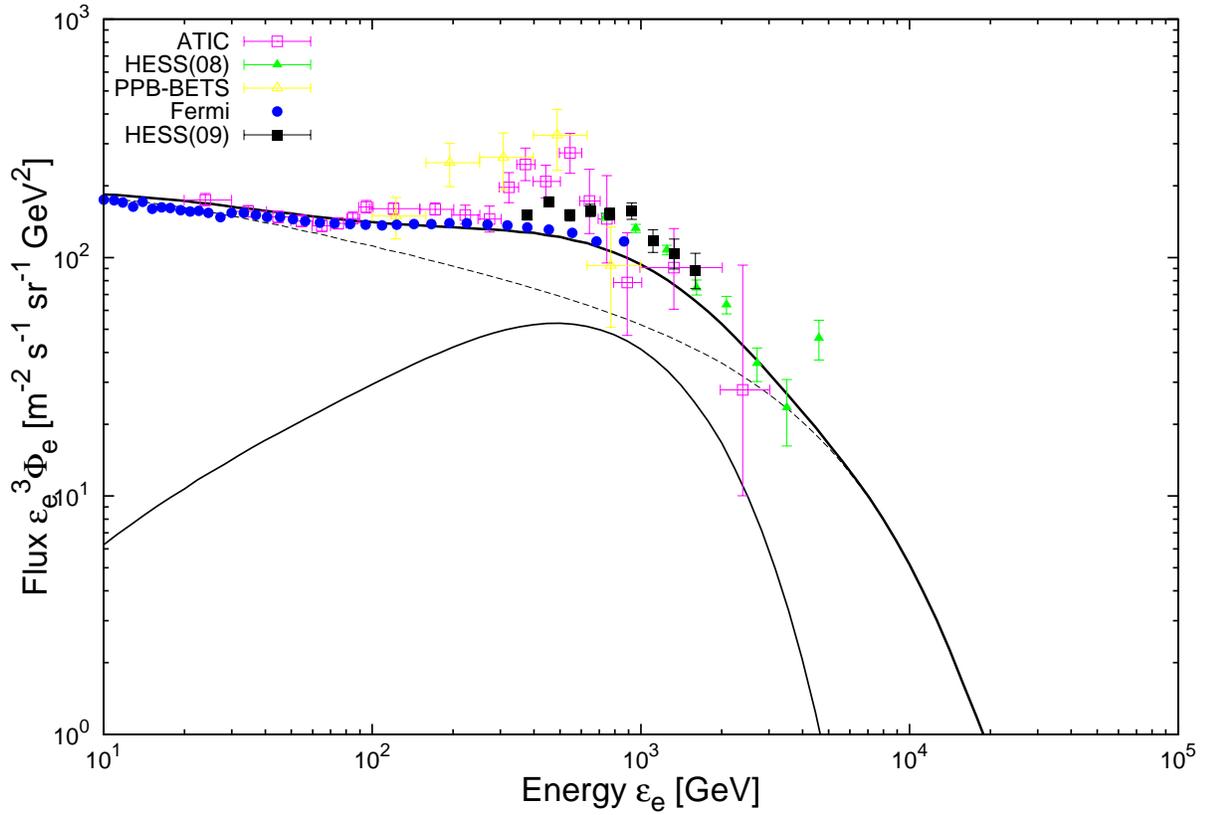}
 \end{center}
 \caption{Electron/positron spectrum predicted from MSPs which have the power-law injection function with the cutoff energy $\varepsilon_{e,{\rm cut}}=1$ TeV, the minimum energy $\varepsilon_{e,{\rm min}} = 1$ GeV and index $\alpha=2$ (thin solid line). The thick solid line shows the total spectrum. For other parameters, we use the same values as in the case of Figure 1.} 
 \label{fig:3.2}
\end{figure}

\begin{figure}
 \begin{center}
  \includegraphics[width=160mm]{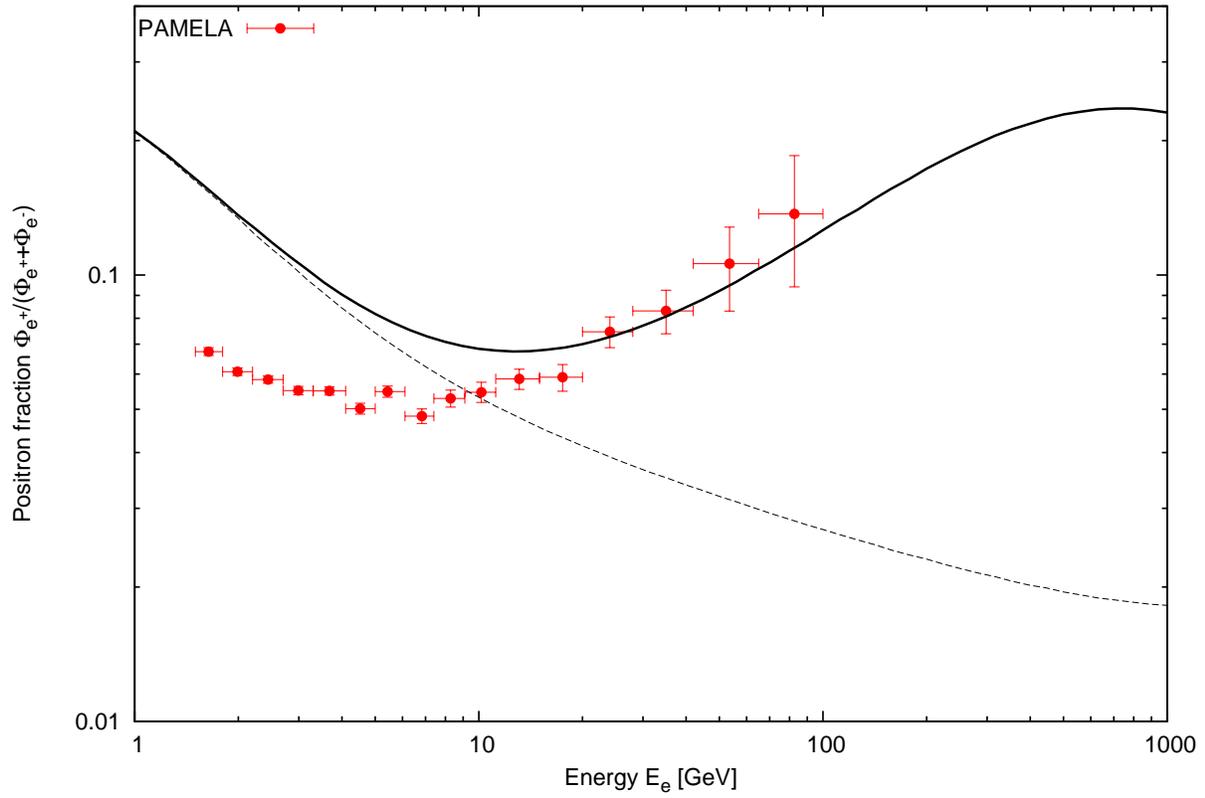}
 \end{center}
 \caption{Total positron fraction resulting from the spectrum (solid line) that have the same parameters as in Figure 2, and the background (dotted line), compared with the PAMELA data as red circles \citep{pamera}. Note that the solar modulation is important below $\sim 10$ GeV.} 
 \label{fig:3.3}
\end{figure}

Next, we also investigate the contribution of MSPs with high pair multiplicity to the observed cosmic ray electrons/ positrons. In this model, we assume that the injection function of these MSPs is power-law distribution with index $\alpha=2$, the cutoff energy $\varepsilon_{e, {\rm cut}}=1$TeV and the minimum energy $\varepsilon_{e, \min}=1$GeV. In this case, the pair multiplicity is $\sim 2000$. Other parameters are the same values as in the case of the pair starved MSPs. We assume that the fraction of MSPs with high multiplicity is 100\%. The results are shown in figure \ref{fig:3.2} for the electron/positron spectrum and in figure \ref{fig:3.3} for the positron fraction. Both figures show that electrons/positrons ejected from MSPs with high multiplicity partially contribute to the excess from background flux observed by PAMELA, HESS and {\it Fermi}. In this energy range, the other sources such as canonical pulsars \citep{S70, AAV95, CCY96, ZC01, Ko04, G07, Bu08, P08, HBS09, YKS09, MCG09, Gr09, KIN10, HGH10} would also contribute to the observed electron/positron spectrum. Note that even if other sources are dominant for the observed excess, the total spectrum added to the contribution of MSPs with high multiplicity does not significantly exceed the observed electron/positron spectrum. 

\section{SUMMARY}

In this paper, we show the possibility that the cosmic ray electrons/positrons from MSPs would significantly contribute to the observed spectrum. Although MSPs have relatively low spin-down luminosity and low birth rate, the lifetime is so long that there are many active MSPs in the vicinity of the Earth. Furthermore, such a long lifetime source continuously injects electrons/positrons after the nebula ceases expanding, so the adiabatic energy losses in a pulsar wind nebula region are negligible. The synchrotron cooling in the nebula is also small, so the high-energy electrons/positrons can escape the nebula without losing much energy. We calculate the diffusive propagation of high-energy electrons/positrons in the ISM taking into account the cooling via synchrotron emissions and inverse Compton scatterings, and predict their spectrum observed at the Earth. 

In the case of the MSPs with multiplicity that is unity, the typical energy of the electrons/positrons produced should be $\sim 50$ TeV based on the assumption of the equipartition in the MSP wind region. Since the long duration of injection make a hard spectrum, the peak is enhanced at 10 - 50 TeV energy range. Even if the fraction of pair starved MSPs is as small as 10\%, this peak would be detectable in the future missions such as CALET and CTA. Although a single young source can make the similar spectral feature in this energy range, in the case of pair starved MSPs the anisotropy of the electron/positron flux would be weaker because a number of sources contribute to it. If this peak is detected, that will be a great impact for on the studies of pulsar radio emission mechanisms because the existence of pair starved MSPs suggests that the radio emission mechanisms should be insensitive to the particle number density down to sub-GJ number density. The detection also suggests that the current outer gap model should be modified because \citet{WH11} suggested that most MSPs locate above the pair death line of the outer gap and the multiplicity is larger than unity. 

We also calculate the electron/positron spectrum from MSPs with high pair multiplicity. We suggest that if multiplicity of these MSPs is the order of $\sim 10^3$, electrons/positrons from them partially contribute to the observed excess of the total spectrum and the positron fraction. 

%

\section*{Acknowledgements}

We thank K. Ioka, K. Kashiyama, T. N. Kato, Y. Kojima, J. Takata and S. J. Tanaka for useful discussions and comments. This work was supported in part by the Grant-in-Aid for Scientific Research from the Japan Society for Promotion of Science (S.K.).


\clearpage

\clearpage

\clearpage

\end{document}